\documentclass[a4paper]{rspublic}%

\newcommand{\prl}{\textit{Phys. Rev. Lett. }}

\newcommand{\pre}{\textit{Phys. Rev. E }}
\newcommand{\ket}[1]{| #1\rangle}                       %
\newcommand{\bra}[1]{\langle #1}                        %
\newcommand{\brac}[1]{\langle #1|}                      %
\newcommand{\etal}{\textit{et al. }}

\newcommand{\De}{\text{\textbf{\textsf{D}}}_{\epsilon}} %quantum coarse 
\newcommand{\cqp}{c_{\epsilon}(q,p)}                    %
\newcommand{\Tqp}{T^{\ }_{qp}}           					\newcommand{\Tqpd}{T_{qp}^{\dagger}}     					         					\newcommand{\equa}[1]{equation~(\ref{#1})}       					\usepackage{graphicx,graphics,wrapfig,rotating}	% Include figure files
\usepackage{dcolumn}
\usepackage{bm,pstricks,amsbsy}
\usepackage{hyperref}

\begin{document}
\title[Irreversibility  in quantum maps with decoherence]{Irreversibility  in quantum maps with decoherence} 
\author[ I. Garc\'ia-Mata, B. Casabone \& D. A. Wisniacki]{Ignacio Garc\'ia-Mata\,${}^{1}$\footnote{Electronic address: garciama@tandar.cnea.gov.ar}, Bernardo Casabone\,${}^{2}$, Diego A. Wisniacki\,${}^{2}$}
\affiliation{
${}^{1}$\,Departamento de F\'isica, Lab. TANDAR, Comisi\'on Nacional de Energ\'ia 
At\'omica, Av. del Libertador 8250, 
C1429BNP  Buenos Aires, Argentina\\
${}^{2}$\,Departamento de F\'isica, FCEyN, UBA, Pabell\'on 1 Ciudad Universitaria, C1428EGA Buenos Aires, Argentina
%\\
%{\rm \small \bf 25 March 2010}
}
\label{firstpage}

\maketitle
\begin{abstract}{Quantum echoes, quantum maps, decoherence}
The Bolztmann echo (BE) is a measure of irreversibility and sensitivity to perturbations  for non-isolated systems. 
Recently,  different %universal 
regimes of this quantity were described for chaotic systems. 
There is a perturbative regime where the BE decays with a rate given by the sum of a term depending on the accuracy with which the system is time-reversed and a term depending on the coupling between the system and the environment. In addition,  a parameter independent regime, characterised by the classical Lyapunov exponent, is expected. In this paper we study the behaviour of the BE in hyperbolic maps that are in contact with different environments. We analyse the emergence of the different regimes and show that the behaviour of the decay rate of the BE is strongly dependent on the type of environment.  
\end{abstract}
\section{Introduction} 

In quantum mechanics there is no ``exponential separation'' of initial conditions due to chaotic motion because evolution is -- in principle-- unitary. Peres (1984)  proposed, as an alternative, to study the stability of quantum motion due to perturbations in the Hamiltionian. As a consequence, the Loschmidt Eco (LE) (Peres 1984; Jalabert \& Pastawski 2001; Jacquod  \emph{et al.} 2001; see two reviews: Gorin \etal 2006 and Petitjean \& Jacquod 2009) 
\begin{equation}
\label{eq:LE}
M(t)=\left|\brac{\psi_0}e^{iH_\Sigma t/\hbar}e^{-iH t/\hbar}\ket{\psi_0}\right|^2
\end{equation} 
was introduced with the purpose of characterising the sensitivity and irreversibility arising from the chaotic nature of quantum systems. The parameter $\Sigma$ denotes perturbation strength. Equation~(\ref{eq:LE}) has a dual interpretation. On  the one hand, it can be interpreted as how close a state remains to itself evolving under slightly different Hamiltonians. On the other hand, it measures the sensitivity of a system to imperfect time inversion, i.e. evolve forward in time under $H$ and then invert time and evolve backward with $H_\Sigma$ (supposing that the time inversion operation is not perfect). 

Depending on the nature of the underlying dynamics, the LE can exhibit qualitatively different behaviour and it thus can be 
used to characterise quantum  chaotic systems. Moreover a number of time-reversal experiments have been performed (Hahn 
1950; Rhim \etal 1970; Zhang \etal 1992; Pastawski \etal 2000), and therein lies the importance of the LE.  
In addition,  the LE (which in quantum information is known as \emph{fidelity}) can be efficiently measured in quantum information systems, i.e. its measurement scales only polynomially with the system size (Emerson \textit{et al.} 2002).

An important fact to remark  is that quantum systems cannot  be isolated easily. Most of the times, there is an environment acting upon the system. This interaction is  most likely unknown and its effects may be uncontrollable. The Boltzmann echo (BE) was introduced (Petitjean \& Jacquod 2006) as a generalisation of the LE to take into account the fact that quantum systems are not isolated.
The idea is to consider the evolution of a system $s$ with a Hamiltonian $H_s$ which is coupled to a an environment $e$ whose evolution is given by $H_e$. We suppose the evolution of the environment $e$ 
is unknown and are therefore uncontrollable, so we trace out the environment degrees of freedom. 
Given a separable  initial state, such as $\rho_0=\rho^{(s)}_0 \otimes \rho^{(e)}_0$, where we take $\rho^{(s)}_0=|\psi_0 \rangle \langle \psi_0 |$,
the BE is defined as the partial fidelity
\begin{equation}
M_B(t) =\Big\langle \langle \psi_0 | {\rm Tr}_{e}[e ^{-iH_bt/\hbar} e^{-iH_ft/\hbar} \rho_0   e^{iH_ft/\hbar} e ^{iH_bt\hbar}]|\psi_0 \rangle\Big\rangle,
\label{eq:ecobol}
\end{equation}
where $H_b$ and $H_f$ are given by
\begin{align}
 H_f&=H_s \otimes \mathbb{I}_e + \mathbb{I}_s\otimes H_e + U_f  \label{eq:Hf}\\
 H_b&=-(H_s + \Sigma_s) \otimes \mathbb{I}_e + \mathbb{I}_s\otimes -(H_e + \Sigma_e) + U_b\label{eq:Hb},
\end{align}
and represent the forward and backward Hamiltonian respectively. Equation~\eqref{eq:ecobol} can be explained as follows. First take an initial state $\rho_0$ and evolve it forward up to time time $t$ with Hamiltonian $H_f$. Then, invert time evolution and evolve with Hamiltonian $H_b$. The imperfection in the inverting process is represented by: $\Sigma_s$ for the system, $\Sigma_e$ for the environment. The terms $U_f,\, U_b$ represent forward and backward interaction between system and environment (for simplicity throughout this work we consider $U_f=-U_b$). 
Finally, the evolution of the system and the BE is obtained by performing a partial trace over the environment degrees of freedom and computing the overlap.
Tracing out the environment makes the effective evolution of the system non-unitary producing decoherence (Zurek 2003). 
An average over initial states of the environment $\rho_0^{(e)}$ is necessary (represented with big brackets in equation \eqref{eq:ecobol}) because we have no control over its degrees of freedom.

In the work of Petitjean \&  Jacquod (2006) the BE was studied semiclassically for two interacting -- classically chaotic -- sub-systems. One of them was used as system and the other as an environment. They found three different regimes for the BE as function of time: parabolic or Gaussian for very short times;  exponential for intermediate, followed by a saturation depending on the effective Hilbert space size. Here we focus on the exponential regime and specifically on the dependence of the decay rate on the perturbation and environment parameters.  
The authors show   (Petitjean \&  Jacquod 2006, see also Petitjean \&  Jacquod 2009) that in the Fermi golden rule (FGR) regime (small perturbation and weak coupling with the environment) the decay rate of the BE results from 
the sum of the decay rates of the LE  due to imperfect time inversion (by definition the BE in the limit no decoherence is just the LE), and the contribution due to the interaction $U_f,\,U_b$ with the environment,
%\begin{equation}
%\label{eq:sumlaw}
 $\Gamma=\Gamma_{\Sigma_s}+\Gamma_f+\Gamma_b$.
%\end{equation}
Henceforth, we call this the \emph{sum law}.
Moreover, for chaotic systems they find that in the limit of strong environment coupling or large perturbation the decay rate is perturbation independent and is given by  the classical Lyapunov exponent. 

In the present contribution, we study the BE  for quantum maps on the torus that are classically chaotic. 
Quantum maps are very simple models that have all the main features of chaotic systems and are ideal for numerical studies. 
Our goal is to understand the behaviour of the BE under the action of different environments. For this reason we have computed the decay of the BE for a wide range of the 
parameters that control the perturbation of the system and the interaction with the environment.
We find that a sum law
for the decay rate of the BE exists. It can be expressed as the sum of the decay rates of the LE 
and the purity of the system, but it is
fulfilled only partially, depending on the decoherence model. The  decoherence models that we present  can be written as a convolution with a kernel. It is for the cases where the kernels have polynomially  decaying tails --models with somewhat large correlations in phase space-- when the sum law is best achieved.  
In addition, the oscillations of the decay rate of the LE, found in e.g Wang (2004), Andersen (2006) and Ares \& Wisniacki (2009), are damped completely in the limit of strong decoherence. However, the decoherence (and perturbation) independent decay rate saturation at the classical Lyapunov exponent is not present for all decoherence models.

The paper is organised as follows.  In \S\ref{sec:sys} \ref{subsec:qkm} we describe the  quantum kicked maps on the torus, the systems used for our studies. Then in, \S\ref{sec:sys} \ref{subsec:oqm}, we introduce our model of open maps using translations in phase space and the Kraus operator sum form.  The main part of this contribution is \S\ref{sec:BE} which is devoted to the numerical calculations and presentation of the results.  Finally, in \S\ref{sec:conc} we summarise our work and results.

\section{The system}
	\label{sec:sys}
\subsection{Quantum `kicked' maps}
			\label{subsec:qkm}
Classical maps  generally arise from the discretisation of a differential  equation of the motion -- like e.g. a Poincar\'e surface of section. Nevertheless one can build abstract maps that do not necessarily relate to a differential equation but that can however provide insight into the properties of chaotic dynamics - e.g the baker's map or the cat map.  Like classical maps, quantum maps are usually simple operators with all typical properties of quantum chaotic systems like level spacing statistics.
In addition, there exist efficient quantum algorithms for some quantum maps (e.g. Goergeot \& Shepelyansky 2001; L\'evy \etal 2003). As the Hilbert space grows exponentially with the number of qubits, one could reach the semiclassical limit with a relatively small number of qubits.  
For this reason they are ideal testbeds for current quantum computers in one of their possible uses: quantum simulators (see Schack 2006). 

The systems we consider are quantum maps on the 2-torus. Periodic boundary conditions imply that Hilbert space has finite dimension $N$ and the effective Planck constant is $\hbar=1/2\pi N$. This means that the semiclassical limit is reached as $N\to\infty$. Position and momentum bases are discrete sets $\{q_i=i/N\}_{i=0}^{N-1}$ and $\{p_i=i/N\}_{i=0}^{N-1}$ related by the discrete Fourier transform 
\begin{equation}
\bra{p}\ket{q}=\frac{1}{\sqrt{N}} e^{-(2\pi \ri N) pq}.
\end{equation}

For practical purposes we will consider maps which  can be expressed as two shears --linear or non-linear--
\begin{equation}
\begin{array}{lcl}
p' &=&p-\frac{dV(q)}{dq}\\
q' &=& q-\frac{dT(p')}{dp'}
\end{array}
\  ({\rm mod}\ 1).
\end{equation} 
These maps can be quantised and the associated unitary map can be written as a product of two `kicks'
\begin{equation}
	\label{eq:unit}
U=e^{i2\pi NT(p)}e^{-i2\pi NV(q)}.
\end{equation}
These types of map usually arise from Hamiltonians with  periodic delta-kicks, like the kicked rotator (Chirikov \etal 1988)  or the kicked Harper Hamiltonian (Leboeuf \etal 1990) . One of the advantages of implementing these types of maps numerically is due to the possibility of using the fast Fourier transform.
%%%%%%%%%%%%%%%%%%%%%%%%%%%%%%%%
\subsection{Open quantum maps}
		\label{subsec:oqm}
A system with an evolution given by a map $U$ might interact with another system acting as environment. If the dynamics of the environment cannot be accessed or controlled then the usual procedure is to trace out  the degrees of freedom of the environment. Tracing out the environment 
translates into a loss  information about the evolution, hence the word \emph{open} -- we picture information flowing out of the system. It is this loss of information the cause of decoherence -- and subsequent loss of \emph{quantumness} (Zurek 2003).  
For a Markovian environment and in the weak coupling limit, this is given by  a completely positive--trace preserving map of density matrices into density matrices -- sometimes called superoperator --which generally can be written in Kraus operator sum form (Kraus 1983) 
\begin{equation}		
\rho_t=\sum_i K_i\rho_{t-1} K_i^\dag,
\end{equation}
where trace preservation is assured by $\sum_i K_i^\dag K_i=I$ ($I$ is the identity).\footnote{Throughout this contribution the `time' $t$ is a discrete time variable which implies 
the number of times a map (or a superoperator) has been applied.}
Therefore the action of the environment is coded into the Kraus operators $K_i$, in analogy  with the Lindblad master equation (Lindblad 1979) where the action of the environment is given by  the Lindblad operators. 
Different Kraus operators will give different types of environments. Rather than modelling the environment through the Lindblad operators and solving the master equation, here we directly model the effect of the environment  on the density matrix of the system by
\begin{equation}
\rho_t\stackrel{\rm def}{=}\De(\rho_{t-1})=\sum_{p,q=0}^{N-1}\cqp\Tqp\rho_{t-1}\Tqpd ,
\label{eq:De}
\end{equation}
where $\Tqp$ are the translation operators on the torus,  $\cqp$ is a function of $q$ and $p$ and $\epsilon$ quantifies the strength of the system-environment coupling.  Even though position and momentum operators with canonical commutation rules are not defined on the torus, translations \emph{can} be defined as cyclic shifts  over the bases elements (Schwinger 1960). Since $\Tqp$ are unitary, trace preservation in equation \equa{eq:De} requires that $\sum_{q,p}\cqp=1$. 
The action decoherence superoperator $\De$ introduced by \equa{eq:De} has a simple interpretation: it implements every possible translation in phase space with probability $\cqp$. This effect is clear in the Wigner function representation. Let $W(q,p)$ be the discrete Wigner function (see e.g. Bianucci \etal 2002) of a density matrix $\rho$ then equation \equa{eq:De} can be re-written as a convolution with $\cqp$
\begin{equation}
  \label{eq:wig}
  W_t(Q,P)=\sum_{q,p}\cqp W_{t-1}(Q-q,p-P).
\end{equation}
This is an incoherent sum of slightly displaced Wigner functions. Any fast oscillating term present in the  state represented by $W(q,p)$ will be eventually washed out, depending on the form of $\cqp$.

For simplicity we suppose that the complete evolution of the quantum map and the decoherent part take place in two steps: first the unitary map $U$  followed by the decoherence term of \equa{eq:De} 
\begin{equation}
\rho_t=\De (U\rho_{t-1} U^\dag).
\label{eq:completeev}
\end{equation}
This is an approximation that works exactly in some cases, e.g. a billiard that has elastic collisions on the walls and diffusion  in the free evolution between collisions. This kind of two-step model has been used to study quantum to classical correspondence and the emergence of classical properties from the quantum dynamics (Nonnenmacher 2003; Garc\'ia-Mata \& Saraceno 2004)

{The effect of decoherence can be characterized by using the purity
\begin{equation}
P(t)={\rm tr}(\rho^2_t),
\end{equation}
were $\rho_t$ is the reduced density matrix of the system.
The purity measures the relative weight of the non-diagonal matrix elements. It is a basis dependent measure that can be used to quantify  the amount entanglement between two parties. If  $P(t)=1$, it means that the global system can be factorized into two separate systems and there is no entanglement. On the contrary, if the purity of the reduced density matrix is minimum (completely mixed state), then the entanglement is maximal. In the case of an $N$ dimensional system  $P(t)=1/N$ for a completely mixed state (maximally entangled with the environment).
As a function of time, after an initial short transient, the purity decays exponentially. For long times it saturates to a minimum value given by
$\hbar/(2\pi)$.}

\section{Numerical results}
	\label{sec:BE}
For our numerical calculations we use the cat map perturbed  in position and momentum  with a smooth  non-linear shear 
\begin{equation}
\begin{array}{lcl}
p'&=&p+a\,q - 2\pi k\sin(2\pi q)\\
q'&=&q+b\,p'- 2\pi k\sin(2\pi p')\\
\end{array}
\  ({\rm mod}\ 1),\label{eq:pcat}
\end{equation}
with $a,b$ integers.
This map is uniformly hyperbolic and fully chaotic. Fo $k\ll 1$ the largest Lyapunov exponent given by 
$\lambda\approx \ln((2+ab+\sqrt{ab(4+ab)})/2)/2$. 
According to \equa{eq:unit} the quantum version of \equa{eq:pcat} is 
\begin{equation}
U_k=e^{2\pi\ri (-P^2/(2N) - k \cos(2 \pi P/N))}\,e^{2\pi\ri (Q^2/(2N) +k \cos(2 \pi Q/N))},
\end{equation}
where $P,\, Q=0,\,\ldots,\, N-1$. All the arithmetic peculiarities of the cat map, which account for the non-generic spectral statistics are destroyed for $k\ne 0$ ( Basilio de Matos \& Ozorio de Almeida 1995;  Keating \& Mezzadri 2000). We can rewrite 
equation \eqref{eq:ecobol} for the  BE  for our open map as the overlap between two states evolving forward in time -- with slightly different maps plus decoherence --  as
\begin{equation}
M_B(t) =  {\rm Tr [ \bar{\rho_{t}}\rho_t]}
\label{eq:mbfinal}
\end{equation}
where 
\begin{align}
\rho_t&= \De (U_k  \rho_{t-1}  U_k^\dag) ,  \\
\bar{\rho}_t&=\De (U_{k'}   \rho_{t-1}  U^\dag_{k'}),
\end{align}
where
 $k,\,k'$ are the perturbation strength of the cat map. We measure the perturbation of one map with respect to the other by the parameter
\begin{equation}
\Sigma\equiv|k'-k|.
\end{equation}
%Equation \eqref{eq:mbfinal}  has two limit cases: In the limit $\epsilon\to 0$ (no decoherence) it gives the LE; in 
%the limit $\Sigma\to 0$  (perfect time-reversal operation) it gives the purity.
For a chaotic system, after an initial transient the BE decays exponentially (Petitjean \& Jacquod 2009). 
Here  we focus on the decay rate  $\Gamma$  as a function of  $\Sigma$ and $\epsilon$ for the exponential decay regime. In the limit $\epsilon\to 0$ we have $\Gamma=\Gamma_\Sigma$, where $\Gamma_\Sigma$ is the decay rate of the LE. In the limit $\Sigma\to 0$ the BE as defined in equation~\eqref{eq:mbfinal}
is equal to the purity, so decay rate is given by the decay rate of the purity $\Gamma_\epsilon$.

We explore the behaviour of $\Gamma$ for three decoherence models and a wide range of values of $\epsilon$ and $\Sigma$. We analyse the parameter domain of validity of the sum law (now $\Gamma=\Gamma_\Sigma+\Gamma_\epsilon$) for these models.
The different models of decoherence we consider are implemented simply by changing the coefficients $\cqp$ in \equa{eq:De}.
 Like for the LE, to extract the decay rate $\Gamma$ an average over an ensemble of initial states needs to be performed. For the averages we used $n_s=10$ randomly chosen coherent states.
%%%%%%%%%%%%%%%%%%%%%%%%%%%%%%%%%%%
\begin{figure}[htbp]
\begin{center}
\includegraphics[width=0.8\linewidth]{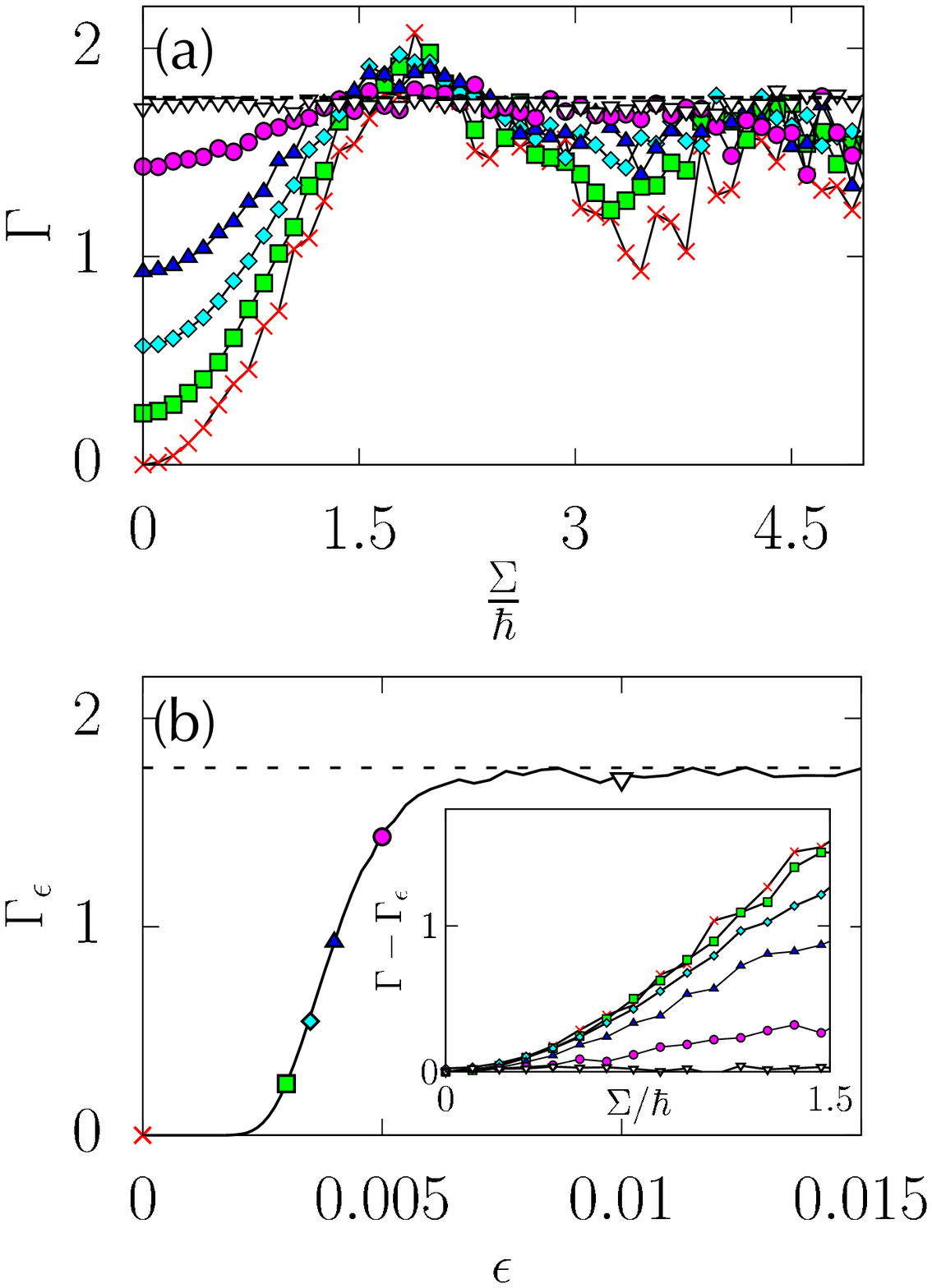}
\caption{
\textbf{(a)} Decay rate $\Gamma$ of the LB as a function of the  rescaled strength of the perturbation $\Sigma/\hbar$ for a GDM environment. The map is the quantum version of the perturbed cat [\equa{eq:pcat}] with $a=b=2$.
 Averages were done over $n_s=10$ initial states.
 Other parameters are: $k=0.001$, $N=800$. 
({$\boldsymbol{\times}$}) $\epsilon=0$ (LE), 
({$\boldsymbol{\square$}}) $\epsilon=0.003$, 
($\boldsymbol{\lozenge}$) $\epsilon=0.0035$, 
({$\boldsymbol{\vartriangle}$}) $\epsilon=0.004$, 
({\large $\boldsymbol{\circ}$}) $\epsilon=0.005$, 
({\large $\boldsymbol{\triangledown}$}) $\epsilon=0.01$. 
The horizontal dashed (in \textbf{(a)} and \textbf{(b)}) lines correspond to the Lyapunov exponents of the corresponding map $\lambda=\ln [3+2\sqrt{2}]\approx 1.76275$; 
\textbf{(b)} The decay decay rate $\Gamma_{\epsilon}$ of the purity as a function of the perturbation parameter $\epsilon$. The points correspond to  the initial values of the curves in \textbf{(a)}. \textbf{(inset)} Decay rate $\Gamma - \Gamma_{\epsilon}$   as a function of the  rescaled strength of the perturbation $\Sigma/\hbar$.
\label{fig1} 
}
\end{center}
\end{figure}
%%%%%%%%%%%%%%%%%%%%%%%%%%%%%%%%%%%%%%
\subsection{Gaussian diffusion}
The first model we have considered was introduced in the work of Garc\'ia-Mata \etal (2003) to model diffusion in a quantum map. We take a periodic sum of Gaussians -- to fit the boundary conditions of the 2-torus -- 
\begin{equation}
\label{cqp_GDM}
\cqp=\frac{1}{A} \sum_{j,k=-x}^x\exp\left[-\frac{(q-jN)^2+(p-kN)^2}{2\left(\frac{\epsilon N}{2\pi}\right)^2}\right],
\end{equation}
where  $x$ is large enough (typically of order 10-15) so that the tails of the furthermost Gaussians can be neglected and $A$ is the normalisation factor ($q,p=0,\ldots, N-1$). We call this model Gaussian diffusion model (GDM). 
The GDM can be interpreted as a smoothing or coarse graining of the unitary evolution:  with Gaussian weight the state is displaced all over a region of size of order $\epsilon$.
As a consequence, the interference terms get washed out, while the remaining classical part is diffused. As stated before, in the continuous limit 
 \equa{eq:wig} is a convolution of the Wigner function with a kernel $\cqp$. For the GDM it can be related to the solution of the heat equation with diffusion constant given by $(\epsilon/2\pi)^2$ (Zurek \&  Paz 1994; Strunz \&  Percival 1998; Carvalho \emph{et al.} 2004; Wisniacki \& Toscano 2009).

In figure \ref{fig1}(a) we show the decay rate $\Gamma$ of the BE as function of perturbation parameter $\Sigma$ for the perturbed cat map $a=b=2$, $N=800$ and $k=0.01$ in the presence of GDM for distinct values of $\epsilon$.  The Lyapunov exponent $\lambda=\ln[3+2\sqrt{2}]$ is marked by a dashed line. 
For $\epsilon=0$ (red $\times$ symbol) we recover the decay rate of the LE: for small $\Sigma$ we get the characteristic quadratic behaviour for small perturbation -- Fermi golden rule regime; for larger values of $\Sigma$ we get a non-universal -- perturbation dependent -- oscillatory behaviour which has also been observed in the work of Wang \etal (2004), Ares \& Wisniacki (2009) and Casabone \etal 2010.
As $\epsilon$ increases, the initial $\Gamma$ value tends to increase (giving the characteristic exponential decay of the  purity  rate due to decoherence) while the amplitude of the oscillations seem to decrease approaching the value of the classical Lyapunov exponent.
% according to  predictions.
 In figure~\ref{fig1} (b)  the decay rate of the purity $\Gamma_\epsilon$, which corresponds to the BE for $\Sigma=0$).
The coloured points correspond to the curves -- for different $\epsilon$ values -- in figure~\ref{fig1} (a). For the GDM we observe saturation of $\Gamma_\epsilon$ at $\lambda$ as is expected.  In the inset we assess the sum law $\Gamma\sim\Gamma_\epsilon+\Gamma_\Sigma$ for the BE. There we plot $\Gamma-\Gamma_\epsilon$ as a function of $\Sigma/\hbar$: 
the expected behaviour -- all curves collapsing into the one corresponding to $\Gamma_\Sigma$ -- is only observed  for values of $\epsilon\lesssim 0.0035$ corresponding to $\Gamma_\epsilon\lesssim 0.5$. For $\epsilon=0.0035$ ($\boldsymbol\lozenge$ symbols) we see that the sum law breaks up around $\Sigma/\hbar\approx 0.75$. For $\epsilon\gtrsim 0.0035$ the sum law is no longer valid. 
%%~~~~~~~~~~~~~~~~~~~~~~~~~~~~~~~~~~~~~~~~~~~~~
\subsection{Generalised depolarising channel}
The next environment  model that we considered is  the generalised depolarising channel (DC). Although -- as we shall see -- in phase space it  is somehow an extremely non-local noise, its importance lies in that it is one of simplest and best known noise channels in quantum information formalism (Nielsen \&  Chuang 2000). The action of  the DC for one qubit ($N=2$) is simple: with probability $(1-\epsilon)$ it does nothing, and with probability $\epsilon$ it `depolarises' it, meaning that it leaves it in a completely mixed state. This is done by  applying every possible Pauli matrix on the state.  For an $N$ dimensional  system, and a torus phase space it can be generalised as follows (Aolita  \emph{et al.} 2004)
\begin{equation}
\De^{\rm DC}=(1-\epsilon)\rho+\frac{\epsilon}{N^2}\sum_{q,p\ne 0}\Tqp\rho\Tqpd
\label{eq:dc}
\end{equation}
that is, with probability $(1- \epsilon)$ it leaves the state unchanged, while with probability $\epsilon$ it applies every possible translation in phase space, with equal weight $\epsilon/N^2$. So contrary to the GDM where the incoherent sum over displaced states took place between states lying effectively close -- due to the Gaussian weight--, for the DC the incoherent sum is over all states, close or apart. It is in this sense that  we say this model is  highly non-local.
%%%%%%%%%%%%%%%%%%%%%%%
\begin{figure}[htbp]
\begin{center}
\includegraphics[width=0.75\linewidth]{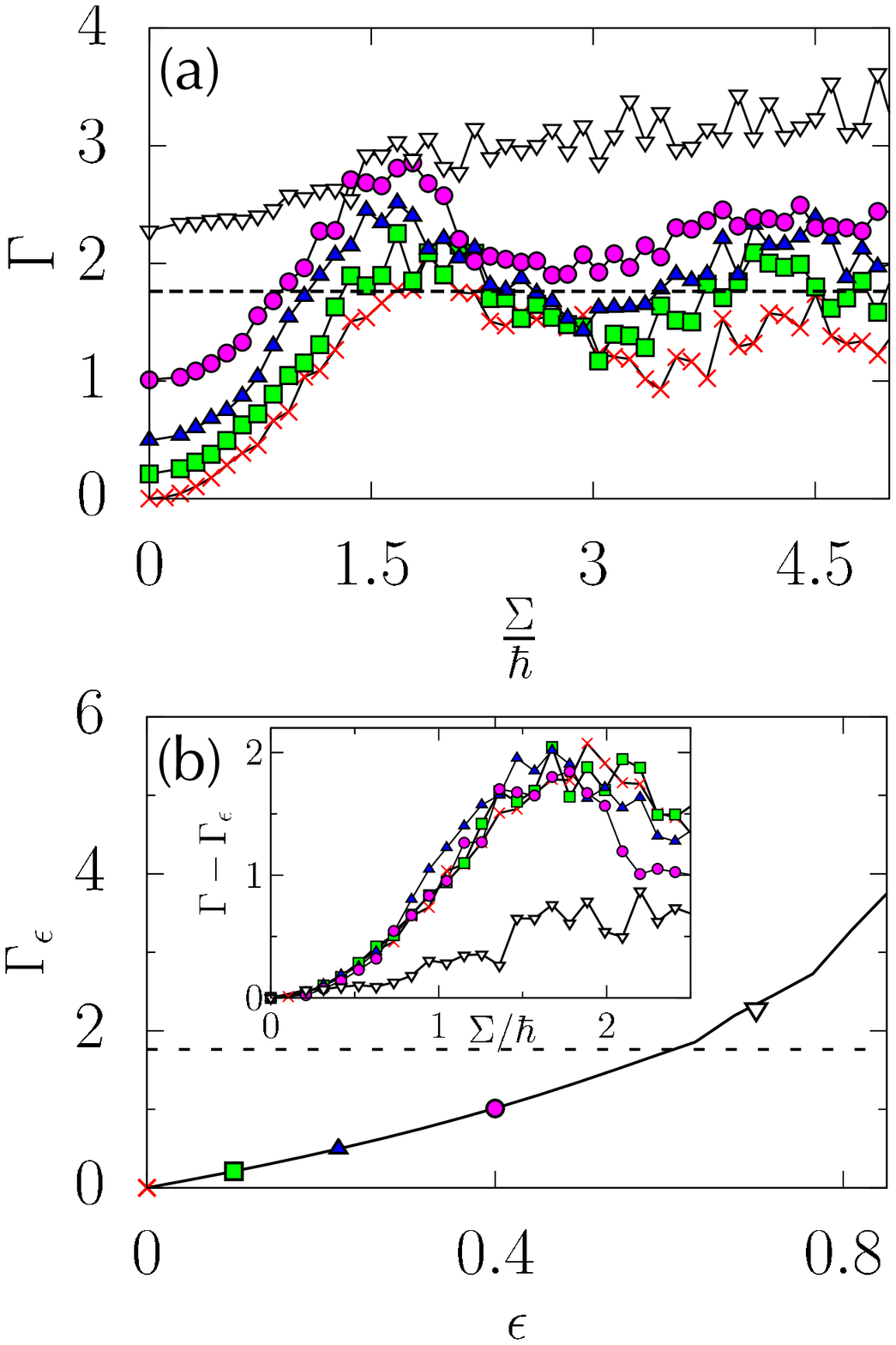}
\caption{
\textbf{(a)} Decay rate $\Gamma$ of the LB as a function of the  rescaled strength of the perturbation $\Sigma/\hbar$ for a DC environment. The map is the quantum version of the perturbed cat [\equa{eq:pcat}] with $a=b=2$.  Averages were done over $n_s=10$ initial states. Other parameters are: $k=0.001$, $N=800$. 
({\large $\boldsymbol{\times}$}) $\epsilon=0$ (LE), 
({ $\boldsymbol\square$}) $\epsilon=0.1$, 
({\large $\boldsymbol\vartriangle$}) $\epsilon=0.22$, 
({\large $\boldsymbol\circ$}) $\epsilon=0.40$,
({\large $\boldsymbol\triangledown$}) $\epsilon=0.7$.
The horizontal dashed (in \textbf{(a)} and \textbf{(b)}) lines correspond to the Lyapunov exponents of the corresponding maps $\lambda=\ln [3+2\sqrt{2}]\approx 1.76275$; 
\textbf{(b)} The decay rate $\Gamma_{\epsilon}$ of the purity as a  function of the perturbation parameter $\epsilon$. The points correspond to  the initial values of the curves in \textbf{(a)}
\textbf{(inset)} Decay rate $\Gamma - \Gamma_{\epsilon}$   as a function of the  rescaled strength of the perturbation $\Sigma/\hbar$.  
\label{fig2} }
\end{center}
\end{figure}
%%%%%%%%%%%%%%%%%%%%%%%%%%%

%%%%%%%%%%%%%%%%%%%%%%%%%%%%%%%%%%%%%
\begin{figure}[htbp]
\begin{center}
\includegraphics[width=0.75\linewidth]{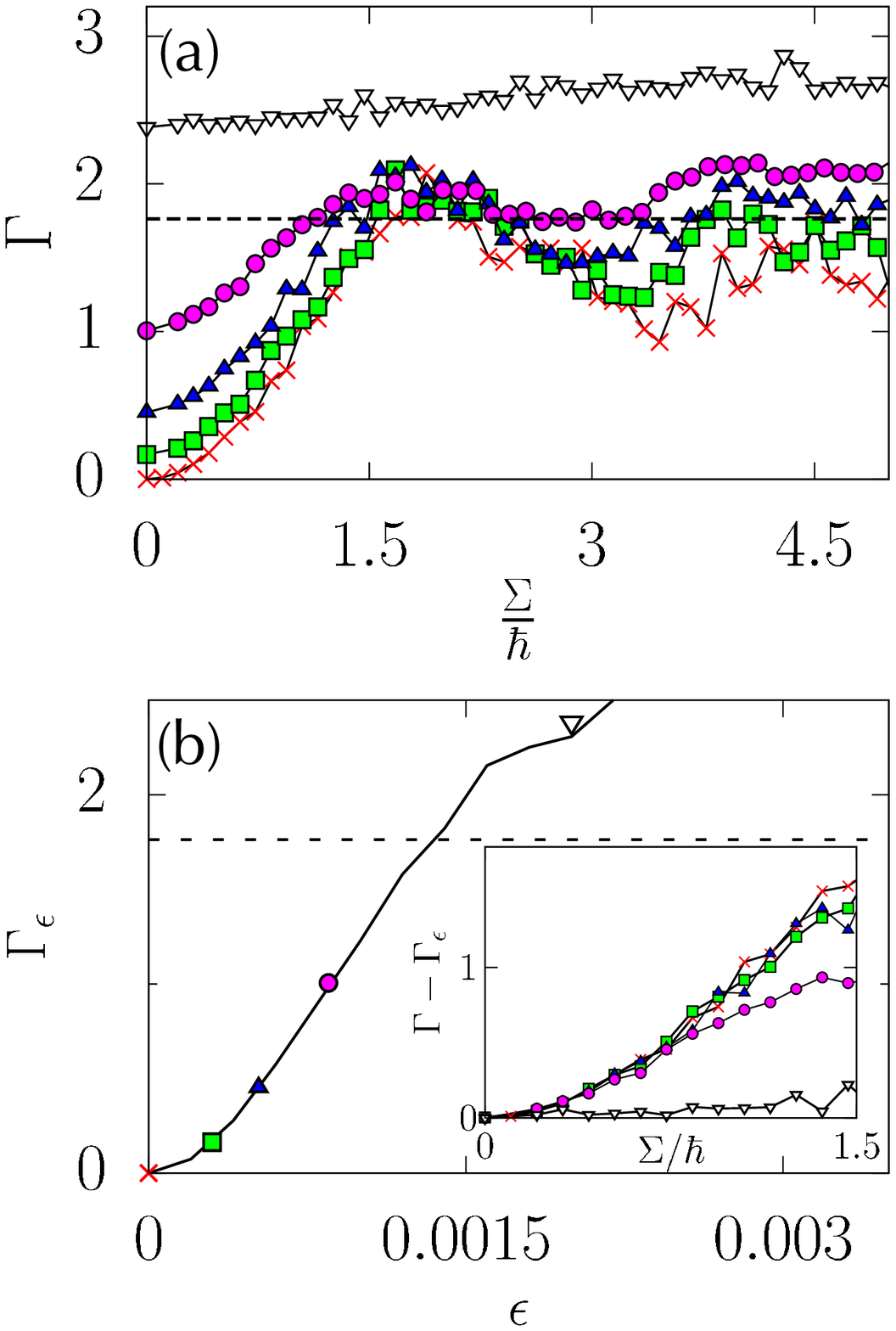}
\caption{
\textbf{(a)} Decay rate $\Gamma$ of the LB as a function of the  rescaled strength of the perturbation $\Sigma/\hbar$ for a LDM environment. The map is the quantum version of the perturbed cat [\equa{eq:pcat}] with $a=b=2$. Averages were done over $n_s=10$ initial states
Other parameters are: $k=0.001$, $N=800$. 
({\large $\boldsymbol\times$}) $\epsilon=0$ (LE), 
({ $\boldsymbol\square$}) $\epsilon=0.001$, 
({\large $\boldsymbol\vartriangle$}) $\epsilon=0.002$, 
({\large $\boldsymbol\circ$}) $\epsilon=0.005$,
({\large $\boldsymbol\triangledown$}) $\epsilon=0.01$.
The horizontal dashed (in \textbf{(a)} and \textbf{(b)}) lines correspond to the Lyapunov exponents of the corresponding map $\lambda=\ln [3+2\sqrt{2}]\approx 1.76275$; 
\textbf{(b)} The decay rate $\Gamma_{\epsilon}$ of the purity as a function of the perturbation parameter $\epsilon$. The points correspond to  the initial values of the curves in \textbf{(a)}.
\textbf{(inset)} Decay rate $\Gamma - \Gamma_{\epsilon}$   as a function of the  rescaled strength of the perturbation $\Sigma/\hbar$. 
\label{fig3} 
}
\end{center}
\end{figure}
%%%%%%%%%%%%%%%%%%%%%%%%%%%%%%%%%%%%%

In figure \ref{fig2} (a) we show the BE decay rate $\Gamma$ as function of perturbation parameter $\Sigma$ for the perturbed cat map $a=b=2$, $N=800$ and $k=0.01$ in the presence of DC noise model for distinct values of  $\epsilon$.  Again, here the red line with $\times$ symbols is $\Gamma_\Sigma$ of the LE.  For smaller $\epsilon$ the curves look like essentially the same curve shifted upwards. There is no evident saturation at the Lyapunov exponent.  
For a larger $\epsilon$ the BE oscillations tend to disappear and the growth is somehow linear with no apparent saturation. 
On figure~\ref{fig2} (b) we show the decay rate of the purity $\Gamma_\epsilon$ as a function of $\epsilon$ and the coloured points mark the initial values of the  curves on the top. Initially $\Gamma_\epsilon$ grows linearly. 
As it is expected (Casabone \etal 2010), there is no parameter independent regime for the DC observed, neither for $\Gamma$ nor for $\Gamma_\epsilon$. 
In the inset of figure \ref{fig2} we show the decay rate of  $\Gamma - \Gamma_\epsilon$. We can see the lines collapse to the curve corresponding to $\Gamma_\Sigma$ (red with $\times$ symbols) for the LE 
for a sizeable interval of $\Sigma/\hbar$ and up to  values of $\Gamma_\epsilon\approx 1$.  
From the work by Casabone \etal (2010) we know that the decay rate of purity as a function of $\epsilon$ is $\Gamma_\epsilon=2\epsilon$, for small $\epsilon$. It is simple to show that in the interval of epsilon where this is valid holds the sum law  $\Gamma_\Sigma\approx \Gamma-\Gamma_\epsilon$ also holds.  Here this is true up to values $\epsilon\lesssim 0.4$ (see also figure~2 in Casabone \etal 2010) correspoding to $\Gamma_\epsilon\lesssim 1$.
%%%%%%%%%%%%%%%%%%%%%%%%%%%%%%%%%%%%%%%%%%%%%%%%%%%%%%%%%%%
\subsection{Lorentzian decoherence}
Finally we consider a model which is more local than the DC but which unlike the GDM has polynomially decaying tails for $\cqp$. The motivation for using this model arose in the work by Casabone \etal 2010 when comparing the universalities of the purity and the LE.  We take $\cqp$ a sum of Lorentzians
\begin{equation}
\label{eq:LDM}
\cqp=\frac{1}{\pi A}\sum_{j,k=-x}^x\frac{\frac{\epsilon N}{2\pi}}{\left( \left(\frac{\epsilon N}{2\pi}\right)^2+(q-Nj)^2+(p-Nk)^2\right)}
\end{equation}
with $A$ the proper normalisation for $\sum_{q,p}\cqp=1$ and $q,p=0,\ldots, N-1$. The sum is done to account for the periodicity of the torus (theoretically $x\to\infty$, practically $x$ is an integer much larger than 1). We call this  model Lorentz decoherence model (LDM). Equation
{ \eqref{eq:De} with $\cqp$ given by \equa{eq:LDM} defines a random process with Lorentzian weight that can be related  to superdiffusion by L\'evy flights. } The effect of heavy tails in decoherence is also explored in e.g the work of Schomerus \&  Lutz (2007).

In figure \ref{fig3}(a) we show the  decay rate $\Gamma$ of the BE as function of perturbation parameter $\Sigma$ for the perturbed cat map $a=b=2$, $N=800$ and $k=0.01$ in the presence of LDM for different $\epsilon$ values.  Again we see that for small $\epsilon$ the curves look like a shift of one another -- although less so than for the DC model-- and then for large values of $\epsilon$ the oscillations are destroyed and the growth of $\Gamma$  is linear, like for the DC. 
On  figure~\ref{fig3}(b) the decay rate $\Gamma_\epsilon$ of the purity with the initial points of the curves on the top superimposed. The initial growth of $\Gamma_\epsilon$ is quadratic with $\epsilon$ as was shown in Casabone \etal (2010). It can also be clearly observed that in neither figure there is a parameter independent --Lyapunov -- regime.
In the inset of figure \ref{fig3} we show the decay rate  $\Gamma - \Gamma_\epsilon$. The  sum law  $\Gamma_\Sigma\sim\Gamma - \Gamma_\epsilon$ holds for an interval of $\Sigma/\hbar$  of up to $\Sigma/\hbar\approx 1.5$ (similar to the DC case) but it seems to break up a little bit earlier in the values of $\Gamma_\epsilon$. Notice that in the inset of figure~\ref{fig3}(b), the line corresponding to the circles ($\Gamma_\epsilon\approx 1$) separates from the others at $\Sigma/\hbar\approx 0.75$.

\section{Conclusions}
	\label{sec:conc} 	
Summarising we have studied the BE for quantum chaotic maps with three different types of decoherence.  The BE complements the original idea of the LE in that it considers the presence of an environment yielding it appropriate for the understanding realistic experiments. 
We have done extensive numerical calculations for a wide range of values of the perturbation of the map and the strength of the decoherence superoperator and we have focused on the decay rate of the BE in the regime where it decays exponentially.
Other than providing a `visual landscape' of the decay rate $\Gamma$ of the BE our calculations  enable a qualitative and quantitative analysis of the universal regimes found in the literature. We found that the more realistic diffusion model (GDM) correctly retrieves the Lyapunov behaviour for large enough values of $\epsilon$.  However, for this same case the sum law $\Gamma\approx\Gamma_\Sigma+\Gamma_\epsilon$ breaks up  for relatively small values of $\epsilon$. We infer that this problem is related to the geometry of phase space (similar non-universal behaviour is found for the purity in the work of Casabone \etal 2010). On the contrary, the two other cases considered satisfy the sum law rather well. These two models have in common the slow decaying tails of the kernel $\cqp$, which means that the decoherence model acts non-locally in phase space. Furthermore these two models fail to exhibit the parameter independent Lyapunov regime.

We have used quantum maps as generic chaotic systems and three very different decoherence models. We can thus conclude that non-generic behaviour is to be expected in echo experiments with arbitrary types environment.

\begin{acknowledgements}
The authors acknowledge financial support from CONICET (PIP-6137), UBACyT (X237) and ANPCyT.  D.A.W.  and I. G.-M. are researchers of CONICET.
\end{acknowledgements}

%\bibliographystyle{unsrt}
%\bibliography{/media/DATA/ber/tesis/tesis2}

\label{lastpage}
%%%
\end{document}